\documentstyle[twocolumn,aps,pra,epsf,epsfig]{revtex}

\begin{document}
\twocolumn[\hsize\textwidth\columnwidth\hsize
\csname@twocolumnfalse%
\endcsname
\draft
\title{Quantum acoustic bremsstrahlung of impurity atoms in a 
Bose--Einstein condensate}
\author{I.E.Mazets}
\address{{\setlength{\baselineskip}{18pt}
Ioffe Physico-Technical Institute, 194021 St.Petersburg, Russia}}

\maketitle

\begin{abstract}
We study the process of scattering of two impurity atoms accompanied by 
generation of an elementary excitation in a surrounding Bose--Einstein 
condensate. This process, unlike the phonon generation by a {\it single} 
impurity atom, has no velocity threshold and can be regarded as a 
quantum acoustic analog of a bremsstrahlung in quantum electrodynamics. 
\\ 
\end{abstract}
\vskip1pc]

Sixty years ago Landau proposed a phenomenologic microscopic description 
of the effect of superfluidity \cite{L41}. He considered a massive body 
moving in a medium with the elementary excitation energy spectrum 
$\epsilon (q)$, 
where {\bf q} is the kinetic momentum of an elementary excitation. 
The energy 
conservation law avoids excitation generation and, therefore, dissipation 
of the kinetic energy of the moving body, if the velocity of the body is 
less than the critical one, 
$v_{cr}=\min _q [\epsilon (q)/q]$. In the case of 
the Bogoliubov tipe spectrum of elementary excitations in a dilute 
Bose--condensed gas \cite{B47}, 
$v_{cr}$ coincides with the speed of sound in such a degenerate quantum 
gas. It has been known, however, that the Landau criterion applies in 
its strict form only to microscopic objects moving through a 
Bose--Einstein condensate (BEC) \cite{tt90}. 
If the size of an object exceeds the 
healing length of the BEC, the object must be regarded as a 
macroscopic one, and its motion can result in generation of vortex pairs 
in the BEC. Reduction, due to effects of the moving object size, 
of the critical velocity with respect to the value given 
by the Landau's theory  has 
been demonstrated in the MIT group experiments \cite{macr}, 
where a laser 
beam has been used as a macroscopic object stirring the BEC. From the 
other hand, the MIT group also performed an experiment \cite{ch0} 
on probing BEC superfluidity by microscopic objects, 
namely, by atoms transferred 
from the BEC to the untrapped hyperfine state by a Raman laser pulse.
A good agreement of the results with the Landau criterion was 
demonstrated. 
  
The probability of an elementary excitation generation by an 
impurity atom in  a dilute BEC has been calculated by Timmermans and 
C\^ot\'e \cite{tc98}. 

The processes described above are not related to  collisions 
between impurity particles in the presence of a BEC. In the present 
paper we try to fill this gap in the theory of ultracold gases. 
We consider a homogeneous BEC of neutral atoms of the first kind at 
zero temperature with a small admixture of atoms of the second kind 
(impurity atoms) 
traveling through the BEC at different velocities. The mass of an 
atom of the $j$th kind is denoted by $m_j$. The number density of 
atoms in the BEC is $n_1$. 

Let us consider two colliding impurity atoms. Before the collision  
their momenta are $\frac 12{\bf P}\pm {\bf p}$, 
respectively, where {\bf P} is the 
center-of-mass momentum and {\bf p} is the momentum of their relative 
motion in the center-of-mass frame of reference. If the collision is 
accompanied by a generation of an elementary excitation with the 
momentum {\bf q}, in the final state the center-of-mass momentum is 
${\bf P}^\prime ={\bf P}-{\bf q}$, and the momentum of relative motion 
${\bf p}^\prime $ is less by absolute magnitude than $p$, unlike 
a case of elastic collision. The Bogoliubov spectrum for elementary 
excitations gives $\epsilon (q)=\sqrt{q^2/(2m_1)[\,q^2/(2m_1)+
2m_1c_s^2]\, }$, where $c_s$ is the speed of sound in the BEC. 
The number density of impurity atoms is assumed to be small enough 
to neglect interaction with them in the expression for the energy spectrum 
of elementary excitations in the BEC. The energy conservation law 
\begin{equation}
\frac {{\bf P}^2}{4m_2}+\frac {{\bf p}^2}{m_2} 
= \frac {({\bf P}-{\bf q})^2}{4m_2}+\frac {{\bf p}^{\prime \, 2}}{m_2} 
+\epsilon (q) 
\label{en1}
\end{equation}
always admits solutions with $q\neq 0$ and the initial velocities 
of the colliding impurity atoms less (and even much less) than the 
critical velocity $v_{cr}=c_s$. 

For the sake of simplicity, we shall consider hereafter the case of 
{\bf P}~=~0. In this case Eq.(\ref{en1}) takes the form 
\begin{equation}
\epsilon (q)+\frac {q^2}{4m_2}+\frac {p^{\prime \, 2}-p^2}{m_2} 
=0.  \label{en2}
\end{equation} 
It is obvious that even if $p\ll m_2c_s$, the generation of a phonon 
with $q\leq p^2/(m_2c_s)$ is possible. 

In Fig.~1 we plot the five diagrams describing the process mentioned 
above. There are three different kinds of vertices corresponding to 
the different multipliers appearing in the transition matrix element 
$\cal{M}({\bf p},\,{\bf p}^\prime )$. The first one is the only kind 
of vertex appearing in the diagrams in Fig.~1\,(a,\,b), where 
it is denoted by small filled circles. It 
corresponds to the multiplier $g_{12}\sqrt{n_1}(u_K-v_K)/\sqrt{V}$. 
Here $V$ is the quantization volume, 
$u_K=\sqrt{\frac {K^2/(2m_1)+m_1c_s^2}{2\epsilon(K)}+\frac 12}$ and 
$v_K=
\sqrt{\frac {K^2/(2m_1)+m_1c_s^2}{2\epsilon(K)}-\frac 12}$ 
are the Bogoliubov 
transformation coefficients \cite{B47}, {\bf K} is the kinetic momentum 
of an elementary excitation, {\bf K}~=~{\bf Q} for a virtual phonon in 
the intermediate state and {\bf K}~=~{\bf q} for the actual phonon in 
the final state. The coupling constant $g_{12}$ describes interaction of 
atoms of the 1st kind with atoms of the 2nd kind and is introduced 
according to the common definition $g_{ij}=2\pi \hbar ^2(m_i+m_j)a_{ij}/
(m_im_j)$, $i,\,j=1,\, 2$, where $a_{ij}$ is the scattering 
length for {\it s}--wave scattering of an atom of the $i$th kind on an 
atom of the $j$th kind. Then in Fig.1\,(c,\,d) we see vertices denoted 
by large filled circles and proportional to $g_{22}/V$. In Fig.1\,(e), 
there is a vertex denoted by an open circle and proportional to 
$g_{11}\sqrt{n_1}[\, 
u_q(u_Qu_{Q^\prime }-u_Qv_{Q^\prime }-u_{Q^\prime }v_Q) -
v_q(v_Qv_{Q^\prime }-u_Qv_{Q^\prime }-u_{Q^\prime }v_Q) \, ]/\sqrt{V}$. 

\begin{figure}[h]
\vspace{-0.9cm}
\begin{center}
\centerline{\psfig{figure=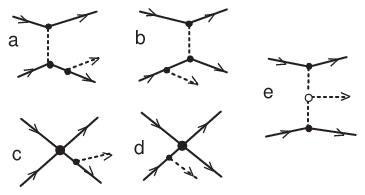,height=5.6cm}}
\end{center}
\vspace{-1.2cm}
\begin{caption}
{The five diagrams describing the process of scattering 
of two impurity atoms 
(solid lines) accompanied by emission of a phonon (dashed line).}
\end{caption}
\label{d5}
\end{figure}

It is worth to note that the diagrams of Fig.~1 (a,\,b) are in certain  
sense analogous of the diagrams describing bremsstrahlung of charged 
particles in quantum electrodynamics \cite{QED}; impurity atoms and 
BEC elementary excitations in our case play role of charged massive 
particles and photons, respectively. However, in our case, the 
attraction potential between two impurity atoms emerging due to virtual 
phonon exchange is a short-range potential. For a pair 
impurity atoms moving with velocities below $c_s$ this potential is 
just of the Yukawa type, and its Fourier transform is \cite{effu}
\begin{equation}
{\cal{U} }_{eff}(Q)=-\frac {g_{12}^2}{g_{11}}\frac 1{1+[Q/(2m_1c_s)]^2}.
\label{Ue}
\end{equation}
Thus, the diagrams of Fig.~1 (a, b) are similar to those of Fig.~2 (c, d) 
where the direct interaction between impurity atoms is short-range, too. 
To make the analogy with quantum electrodynamics closer, we refer to 
the bremsstrahlung process in a collision between a charged 
particle and a neutral atom (see, e.g., the recent paper by Korol and 
co-workers \cite{brsen} and numerous references therein related to 
ordinary as well as to polarizational bremsstrahlung). 
The last diagram shown in  Fig.~1 (e) is distinct from the previous ones. 
It has no analog among third-order processes in quantum electrodynamics 
(only in fifth order a similar process appears, if annihilation of two 
virtual photon and creation of one photon in a final state require 
creation of an electron-positron pair in an intermediate state). Therefore, 
it is reasonable to represent the transition matrix element as the sum 
\begin{equation} 
{\cal{M} }({\bf p},\,{\bf p}^\prime )=
{\cal{M} }^{(1)}({\bf p},\,{\bf p}^\prime )+
{\cal{M} }^{(2)}({\bf p},\,{\bf p}^\prime ),    
\label{MM}
\end{equation} 
where the first term in the right hand side corresponds to the four  
diagrams of Fig.~1 (a --- d) together, and the second term corresponds 
to the diagram shown in Fig.~1 (e). 

We consider here the case of impurity atom velocities less than the 
speed of sound in the BEC. Under this condition elementary excitations 
both in the final and intermediate state belong to the phonon range of 
the Bogoliubov spectrum, and their magnitudes differ significantly: 
$Q\approx \left| {\bf p}-{\bf p}^\prime \right| \gg q$. For the 
diagram of Fig.~1 (e) the momenta of the two virtual phonons are 
practically opposite, since their sum ${\bf Q}+{\bf Q}^\prime ={\bf q}$ 
is relatively small. Therefore the effects of Bose statistics of 
phonons are not pronounced for the diagrams with two elementary 
excitations in the intermediate state [Fig.~1 (b,\,e)], because they 
occupy essentially different modes. 

If the relative velocity of the colliding impurity atoms is small 
compared to the sound velocity then we have also $Q\ll m_2c_s$ and, 
hence, can the processes contributing to 
${\cal{M} }^{(1)}({\bf p},\,{\bf p}^\prime )$ within the simplified 
model of effective contact interaction. I.e., we describe them by the 
diagrams similar to those of Fig. 1 (c,~d) where the vertex 
corresponding to the contact interaction of two impurity atoms is 
substituted by $g_{eff}/V$, where $g_{eff}=g_{22}-g_{12}^2/g_{11}$. 
It is also convenient to define the effective scattering length 
$
a_{eff}=a_{22}-\left( 1+ {m_1}/{m_2}\right) ^2 
{m_2a_{12}^2}/(4m_1a_{11}) , 
$
so that $g_{eff}=4\pi \hbar ^2a_{eff}/m_2$. 

Using the conditions and approximations 
discussed above, we calculate the matrix elements 
\begin{equation} 
{\cal{M} }^{(1)}({\bf p},\, {\bf p}^\prime ) =
-\frac {4g_{eff}g_{12}\sqrt{n_1}}{V^{3/2}c_s\sqrt{2m_1c_sq\, }} , 
\label{M1}
\end{equation}
and
\begin{equation}
{\cal{M} }^{(2)}({\bf p},\, {\bf p}^\prime ) =
\frac {g_{12}^2g_{11}n_1^{3/2}}{2V^{3/2}c_s^2\sqrt{2m_1c_sq\, }
\left|{\bf p}-{\bf p}^\prime \right| }.
\label{M2}
\end{equation}
For evaluation of an energetic denominator 
$(p^2-p^{\prime \, 2})/m_2+q^2/(4m_2)$ appearing in derivation of 
Eq.(\ref{M1}), the energy conservation law [Eq.(\ref{en2})] was used. 
For the further calculations involving 
${\cal{M} }^{(2)}({\bf p},\, {\bf p}^\prime ) $, it is useful to recall 
the well-known expansions \cite{AM} 
$\left| \,{\bf p}\mp {\bf p}^\prime \right|^{-1} =p^{-1}
\sum _{l=0}^\infty (\pm p^\prime /p)^lP_l(\cos \vartheta ),$ where 
$P_l$'s are Legendre polynomials, $\vartheta $ is the angle between 
{\bf p} and ${\bf p}^\prime $, and $p^\prime <p$. One can note that 
the matrix element given by Eq.(\ref{M1}) contributes only to 
$s$-wave scattering while the matrix element given by Eq.(\ref{M2}) 
contributes also to higher angular momentum scattering channels. 

The total cross-section of the process of two impurity atoms collision 
accompanied by a phonon emission can be written, taking into account 
the effects of impurity atoms quantum statistics, as \cite{QM} 
\begin{eqnarray}
\sigma _{tot}&=&\frac 1j
\frac {2\pi }\hbar \int \frac {V\,d^3{\bf p}^\prime }{
(2\pi \hbar )^3} \, \int \frac {V\, d^3{\bf q}}{(2\pi \hbar )^3} 
\frac 12 \left| {\cal{M} }({\bf p},\,{\bf p}^\prime )\pm  \right. 
\nonumber   \\  &   &   
\left.  {\cal{M} }({\bf p},\,-{\bf p}^\prime )\right| ^2 
\delta \left( \epsilon (q)+\frac {q^2}{4m_2}+
\frac {p^{\prime \, 2}-p^2}{m_2}\right)      .
\label{sgt}
\end{eqnarray}
The probability flux density $j=2p/(Vm_2)$ in the denominator 
is necessary for proper 
normalization of the cross-section. The upper sign in Eq.(\ref{sgt}) 
applies if two colliding impurity atoms are identical bosons, and the 
lower sign applies if they are identical fermions. 

After some tedious but straightforward calculations we get the total 
cross-section for a pair of colliding identical bosonic atoms: 
\begin{equation}
\sigma _{tot}(p)=\sigma _s(p)+\sigma _{he}(p)  , \label{stot}
\end{equation}
where the main term 
\begin{eqnarray} 
\sigma _s(p)&=&\frac 1{30}a_{12}^2\sqrt{4\pi n_1a_{11}^3}\, \left( 
1+ \frac {m_2}{m_1}\right) ^2 \left( \frac p{m_2c_s}\right) ^2 
\nonumber  \\   &  &   \times 
\left[ \frac {a_{12}}{a_{11}}\left( 1+\frac {m_1}{m_2}\right) - 
16\frac {a_{eff}}{a_{11}}\frac p{m_2c_s}\right] ^2
\label{sgpdr}
\end{eqnarray}
describes $s$-wave scattering, and the relatively small correction 
\begin{eqnarray} 
\sigma _{he}(p)&=&\kappa _b\,a_{12}^2\sqrt{4\pi n_1a_{11}^3}\, \left( 
1+ \frac {m_2}{m_1}\right) ^2 \left( 1+\frac {m_1}{m_2}\right) ^2 
\nonumber   \\ & &  \times 
\left( \frac p{m_2c_s}\right) ^2\left( \frac {a_{12}}{a_{11}}\right) ^2
\label{sgpp}
\end{eqnarray} 
describes scattering into partial waves with higher even angular 
momenta; 
$\kappa _b=\sum _{\nu =1}^\infty [2(4\nu +1)(4\nu +3)(4\nu +5)]^{-1} 
\approx 0.0023$. 

In the case of two colliding identical fermionic atoms we get  
\begin{eqnarray} 
\sigma _{tot}(p)&=&\kappa _f\,a_{12}^2\sqrt{4\pi n_1a_{11}^3}\, \left( 
1+ \frac {m_2}{m_1}\right) ^2 \left( 1+\frac {m_1}{m_2}\right) ^2 
\nonumber   \\ & &  \times 
\left( \frac p{m_2c_s}\right) ^2\left( \frac {a_{12}}{a_{11}}\right) ^2
, \label{sgff}
\end{eqnarray} 
where $\kappa _f=\sum _{\nu =0}^\infty [2(4\nu +3)(4\nu +5)(4\nu +7)
]^{-1}\approx 0.0060$.  

Since the process of collision of two impurity atoms accompanied by 
a phonon emission appears in higher orders of perturbation theory 
than a simple elastic collisions of two atoms (the cross-section of the 
latter process is of order of scattering length squared), 
the cross-section of quantum acoustic bremsstrahlung given by 
Eqs.(\ref{sgpdr}, \ref{sgpp}, \ref{sgff}) contains an extra 
small multiplier --- the square root of the BEC 
diluteness parameter, $n_1a_{11}^3\ll 1$. An additional 
factor of smallness of this cross-section is the smallness of the  
ratio $p/(m_2c_s)$ for slow collisions. However, it is possible to 
outline an experimentally feasible 
situation where the process of quantum acoustic 
bremsstrahlung cannot be neglected.  

First of all, to suggest a realistic experiment, one needs not to forget 
about  the effects of finite temperature. 
It is practically impossible to get in experiment the temperature lower 
than the chemical potential of a degenerate Bose-gas. At temperatures $T$
of order of or larger than $m_1c_s^2$, quantum acoustic bremsstrahlung 
never can be the main process leading to thermal equilibration between the 
main component (atoms of the 1st kind) and the gas of impurity atoms. 
Certainly, the emission of phonons \cite{tc98} by the supersonic 
fraction of thermal distribution of impurity atoms and the reciprocal 
process recently considered by Montina \cite{mont} dominate. 

Thus we come to conclusion that a process involving bosonic 
stimulation has to be suggested. 

Let us consider two counterpropagating plane matter waves of impurity 
atoms with atomic kinetic momenta equal to $\pm {\bf p}$, correspondingly. 
Each wave contains $N_2\gg 1$ atoms, $n_2=N_2/V$, 
and the total density of impurity atoms is $2n_2$. Additionally, we 
assume that in the BEC a sound wave with the momentum {\bf q} is 
generated, and the phonon number in this  sound wave is denoted $N_{ph}$. 
We assume that both $N_2$ and $N_{ph}$ are so large that any effect 
arising from interaction with the above-condensate fraction at the 
given temperature is negligible compared to the processes described below. 
The momenta of the 
impurity atoms matter wave and the sound wave are chosen to be small, 
so that 
$p< m_2c_s$ and $\epsilon (q)+q^2/(4m_2) < p^2/m_2$. Also generation of 
a phonon with the momentum {\bf q} during the collisions of the two 
impurity atoms with moments $\pm {\bf p}$ is allowed by the energy 
conservation law. 

Since in the final state there are many phonons in the sound wave, the 
quantum acoustic 
bremsstrahlung process with phonon scattering to this particular 
mode is Bose-enhanced by a factor of $N_{ph}+1\approx N_{ph}$, 
and can prevail over other channels, provided that $N_{ph}$ is large 
enough. It leads to amplification of the sound wave. However, there are 
two processes leading to the sound wave attenuation: (i) phonon 
scattering on impurity atoms, and (ii) the process reverse to quantum 
acoustic bremsstrahlung, namely, absorption of a phonon followed by 
an increase of kinetic energy of relative motion of a pair of impurity 
atoms. 

The evolution of number of phonons in the sound wave is governed by the 
equation 
\begin{equation}
\dot {N}_{ph}=\dot{N}_{ph}^{(col)}+\dot{N}_{ph}^{(qb)},       \label{Nn}
\end{equation} 
where the first term in the right hand side corresponds to 
phonon scattering on impurity atoms and the second term corresponds to 
quantum acoustic bremsstrahlung and its reverse process. 
The first term can be written explicitly as follows: 
\begin{equation} 
\dot{N}_{ph}^{(col)}=-8\pi a_{12}^2n_2m_1^{-1}q N_{ph}   .\label{ncol2}
\end{equation}
This expression is exact if $m_2\gg m_1$. 
However, even if $m_1$ and $m_2$ are comparable, Eq.(\ref{ncol2}) 
provides a reasonable approximation, provided that $p\ll m_2c_s$. 

Calculation of the 
rate of change of phonon number in the sound wave due to the 
processes involving a {\it pair} of impurity atoms results 
in the following expression:
\begin{eqnarray}
\dot{N}_{ph}^{(qb)}&=&\frac {N_{ph}N_2^2Vm_2}{2\pi \hbar ^2}
\left[ \frac {p^\prime }2 \overline{
\left| {\cal{M} }({\bf p},\,{\bf p}^\prime  )\pm  
{\cal{M} }({\bf p},\,-{\bf p}^\prime )\right| ^2} \right. 
\nonumber \\ & & \times  
\Theta \left( \frac {p^2-p^{\prime \, 2}}{m_2}-
\epsilon (q)-\frac {q^2}{4m_2}\right) -\nonumber \\  & &    
\left. \frac {p^{\prime \prime }}2 \overline{
\left| {\cal{M} }({\bf p}^{\prime \prime },\,{\bf p} )\pm   
{\cal{M} }({\bf p}^{\prime \prime },\,-{\bf p})\right| ^2} \right]. 
\label{nbs1}
\end{eqnarray}
Here 
$p^\prime =[{p^2-m_2\epsilon (p)-q^2/4}]^{1/2}$ is the momentum of 
a relative motion of an atomic pair after emission of a phonon, and 
$p^{\prime \prime }=[{p^2+m_2\epsilon (p)+q^2/4}]^{1/2}$ is the 
momentum after absorption of a phonon. For the sake of generality,  
we retain here the Heavyside's step function $\Theta $, however, 
by assumption, its argument is positive. Bars over the 
expression denote averaging over the angle between kinetic momenta 
of relative motion of two impurity atoms before and after the 
inelastic collision. Plus or minus sign must be taken in the cases of 
bosonic or fermionic impurity atoms, respectively. 

We write Eq.(\ref{nbs1}) more explicitly for the 
case of bosonic impurity atoms, since in this case 
the rate given by Eq.(\ref{nbs1}) is significantly larger than in 
the case of fermions. 
Taking into account that the isotropic scattering dominates 
for bosonic impurity atoms, we reduce  Eq.(\ref{nbs1})  to 
\begin{eqnarray} 
\dot{N}_{ph}^{(qb)}&=&\frac \pi 8N_{ph}\frac {n_2^2}{n_1}a_{12}^2c_s 
\frac {m_2c_s}p\frac {m_1c_s}q  
\left[ \frac {p^\prime }p\xi (p) \right.  \nonumber \\ & & \times 
\left.   \Theta \left( \frac {p^2}{m_2}
-\epsilon(q)-\frac {q^2}{4m_2}\right)  -
\frac p{p^{\prime \prime }}\xi(p^{\prime \prime})\right] ,
\label{nqb2}
\end{eqnarray}
where $\xi (p)=\left( 
\frac {a_{12}}{a_{11}}-16\frac {a_{eff}p}{a_{11}m_2c_s}\right) ^2$. 
The net effect of the two processes contributing to Eq.(\ref{nqb2}) is 
always sound wave attenuation. 

As an example, consider the case when $m_2\approx m_1\equiv m$, 
all the scattering lengths $a_{ij}\approx a_{12}$ 
(hence, $a_{eff}\ll a_{12}$), 
$p/(mc_s)\approx 0.3$, and $q/(mc_s)\approx 0.07$. If, e.g., we set 
$p=0.3\,mc_s$ and $q=0.07\,mc_s$, Eq.(\ref{Nn}) gives 
$$
\dot{N}_{ph}=-8\pi n_2a^2_{12}c_sN_{ph}\left( 0.07+0.21\,n_2/n_1\right) .  
$$
Thus increase of sound wave dissipation rate by few dozens percents 
seems experimentally feasible. 

This work is supported by the Nederlandse Organisatie
voor Wetenschappelijk Onderzoek, project NWO-047-009.010. One 
of the authors, I.E.M., also acknowledges the support from the
Russian Foundation for Basic Research, grant No. 02--02--17686, and 
the Ministry of Education of Russia, grant No. UR.01.01.040. 
I.E.M. is grateful to Prof. D.A. Varshalovich and Prof. G.V. Shlyapnikov 
for helpful discussions.

\end{document}